\documentclass[conference]{IEEEtran}
\usepackage{epsfig,rotating,setspace,latexsym,amsmath,epsf,amssymb,bm,theorem}
\usepackage{cite}
\usepackage{graphicx}

\begin{document}

\title{ Active Status Update Packet Drop Control in an Energy Harvesting Node \vspace{-0.1in}}

\author{\IEEEauthorblockN{Parisa Rafiee \qquad Omur Ozel}
\IEEEauthorblockA{Department of Electrical and Computer Engineering\\
George Washington University, Washington DC 20052\\
{\it rafiee@gwu.edu \qquad ozel@gwu.edu}}\vspace{-0.2in}} 

\maketitle

\begin{abstract}
This paper considers an energy harvesting sensor node with battery size $B_{max}$ that recharges its battery through an incremental energy harvesting process and receives updates from a single information source in slotted time. The node actively decides to power down (OFF) or up (ON) the communication circuitry for a portion of its operation time in order to maintain energy efficiency. Update packets arriving in ON (OFF) periods are received (discarded). A deterministic energy cost per time is paid during ON periods. The power down decision can be in partial or full nature, yielding various options for deciding ON-OFF intervals. We develop age-threshold based power ON-OFF schemes to minimize age of information at the node subject to energy harvesting constraints with partial and full power down options for $B_{max}=1$ and $B_{max}=\infty$ cases.   
\end{abstract}

\section {Introduction}

This paper explores \textit{information freshness} in an energy harvesting sensor node that replenishes energy through a random power source and receives updates from a single memoryless information source. Coming from other nodes or directly from the environment, these updates represent a physical phenomenon the node tracks. After receiving the update, the node makes the update available to upper application layers. Our focus is on applications sensitive to freshness of information available to them measured by age of information (AoI) metric. Freshness of information measured by AoI at a monitoring device is critical in Internet of Things (IoT) applications targeting real time operation such as sensor and camera networks for event monitoring \cite{kalor2019minimizing,he2018minimizing} and vehicular networks for monitoring vehicles' states \cite{kaul2011minimizing,Alabbasi2018JointIF,giordani2019framework}.
 
Operating energy harvesting devices requires energy efficiency in addition to energy harvesting constraints. In particular, energy efficiency objective dictates that devices \textit{partially} or \textit{fully} power down communication circuits to replenish battery \cite{joseph2009optimal,niyato2007wireless,fafoutis2011odmac}. For such an energy harvesting device, incoming status updates are discarded when it is powered down; therefore, timeliness could be lost at the expense of energy efficiency. Decisions to turn ON or OFF are taken on the fly, and consequences of potential loss of data in the timeliness of information dissemination have to be minimized to the extent possible. This paper is devoted to the study of actively controlling status update packet drops in energy harvesting sensor nodes for maintaining information freshness. 

Existing works in the literature on information freshness in energy harvesting systems \cite{bacinoglu2015age, yates2015lazy,arafa2017age,wu2017optimal, bacinoglu2017scheduling, bacinoglu2018achieving,feng2018age,arafa2018online,arafa2019using} mainly cover \textit{generate-at-will} policies with a single update source with some exceptions \cite{farazi2018age, farazi2018average}. This type of policy refers to the case when generation of updates is under full control of the transmitter and it does not address the need for packet drop control for energy efficiency. To the best of our knowledge powering down devices for energy efficiency and consequent loss of status updates have not been considered in the literature so far; although status update discarding mechanisms and potential AoI performance improvements have been covered in previous works \cite{costa2016age,kam2018age2,bedewy2017age}. Our work builds upon the pursuit of maintaining energy efficiency through controlling loss of status updates. We also note recent works on \textit{energy efficient} delay minimization in \cite{huang2016delay, nath2018optimum}. In these works, receiver nodes power down under no incoming traffic, and then, wake up for successful reception. Additionally, authors in \cite{maatouk2018age} consider queues with vacations to model sleeping intervals. However, none of these works treat receivers as \textit{active} entities that accept or reject flowing information for energy efficiency. Our current work is performed with a motivation to fill this gap.

In this paper, we consider an energy harvesting sensor node receiving updates from a single information source as depicted in Fig. \ref{model}. The node recharges its battery through a Bernoulli energy harvesting process and saves this energy in a battery of size $B_{max}$. For the update packet to be received, the node's device must be turned ON. Note that a deterministic energy cost per time is paid during ON periods only if sufficient energy is available in the battery. Update packets are discarded during OFF periods. The node actively decides to power down (OFF) the device for a portion of its operation duration to stop expending and to start replenishing energy at the expense of increased AoI due to discarded updates. All operations happen in slotted time. We develop age-threshold based power down schemes and obtain closed form expressions for the resulting average age of information (AoI) at the sensor node. 

The power down decisions can be in partial or full nature. In a partial power down scheme, the node is aware of incoming status update packets causally, and can decide to power up and receive the update. In a full power down scheme, the updates are not available to the receiver during OFF intervals; therefore, receiver has to turn on without the knowledge of presence of an update. Still, both cases are equivalent in terms of energy as no (one unit) energy is expended in OFF (ON) time slots and incoming energy arrivals replenish the battery regardless of the state of the device. These power down schemes represent two extreme options and we address these schemes for $B_{max}=1$ and, $B_{max}=\infty$ cases. 

Our approach allows the energy harvesting sensor node at the receiving end to turn its device ON and OFF judiciously by using age-thresholds. This way, the resulting increase in the average age of information due to packet drops in OFF periods is made small to the extent possible. Our analysis sheds light on relations between energy efficiency and information freshness in energy harvesting nodes.

\section{The Model} 
\label{sect_model}

\begin{figure}[t]
\centerline{\includegraphics[width=0.85\linewidth]{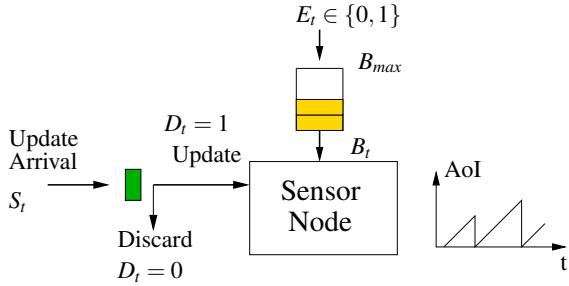}\vspace{-0.1in}}
\caption{System model representing an energy harvesting sensor node receiving incoming updates when it is turned on, and discarding otherwise.}\vspace{-0.55cm}
\label{model}
\end{figure}

We consider an energy harvesting node receiving status update packets from a single information source as in Fig.~\ref{model}. Throughout the document, we consider a slotted system; $t \in \mathbb{Z}_{\geq 0}$ is the time index over which status updates and energy harvests run. We next elucidate details of these models.

\subsection{Information Update Model} 
Information updates at time $t > 0$ are modeled as $S_t \in \{0,1\}$. Each $S_t$ is an independent identically distributed (i.i.d.) binary valued random variable. $S_t=1$ represents the existence of an update in the source at time $t$, and $\lambda=\mbox{Pr}(S_t=1)$ is the corresponding rate of update generation. It is also worthwhile to note that these updates have contents. Our framework is concerned with only the age at the monitoring energy harvesting node. Age of information (AoI) at the monitoring/receiving node is defined as the random sequence
\[  \Delta_t = t - u(t). \]
Here, $u(t)$ is the arrival (generation) time of the most recent successful status update at the receiver. Randomness in $u(t)$ stems from the uncertainty in source arrivals, energy harvests and decisions to power down. The average long-term AoI is
\begin{align*}
\Delta =& \limsup_{n\rightarrow \infty} \frac{\mathbb{E}\left[\sum_{j=1}^{n}Q_j\right]}{\sum_{j=1}^{n}T_j}
=\limsup_{n\rightarrow \infty}  \frac{\mathbb{E}\left[\sum_{j=1}^{n}T_j^2\right]}{2\sum_{j=1}^{n}T_j},
\end{align*}
where $T_i$ is the duration between two updates, and $Q_i=T_i^2/2$ is the total accumulated age between two updates represented by the area\footnote{Due to integer variables, there is an additional $\frac{1}{2}$ term in the age. We exclude it since it does not affect the ordering among different schemes.}. It is crucial to note that if $T_i$ is a random renewal interval independent of other intervals and identically distributed over all $i$, then we can express the average AoI in compact form as
\begin{align*}
\Delta = \frac{\mathbb{E}[T^2]}{2 \mathbb{E}[T]}.
\end{align*}
In the sequel, we will focus on schemes yielding independent $T_i$ and aim to evaluate average AoI in compact form.

\subsection{Energy Model}

Energy arrivals are known causally at the sensor node and are distributed according to an i.i.d.~Bernoulli distribution with parameter $q$, i.e., $\mathbb{P}[E_i=1]=1-\mathbb{P}[E_i=0]=q$. Our model accounts for energy needed to receive status updates and to allow the receiving nodes to \textit{actively} power down their devices and drop some incoming packets \textit{by choice} for energy efficiency. We assume the processing energy needed to keep the transmission circuitry on (active) during a time slot is one unit and is denoted by $D_t \in \{0,1\}$ ($D_t=0$ representing a decision to power down). The battery energy $B_t$ obeys the following rule:
\begin{align*}
B_{t+1} = \min\{B_t - D_t + E_{t+1}, B_{max}\}.
\end{align*}
Here, $B_{max}$ is the battery storage limit and is assumed to be an integer. \textit{Energy causality} constraints due to EH capability are formally stated as $D_t  \leq B_t$ where harvested energy cannot be used before it arrives to the node and is saved in a battery. 

\begin{figure}[t]\hspace{0in}
\includegraphics[width=1.01\linewidth]{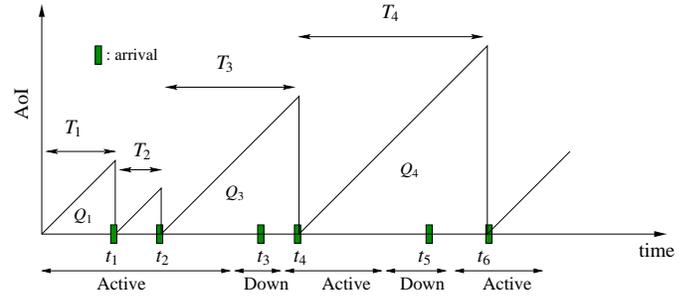}\vspace{-0.05in}
\caption{\label{fig:model11}Update arrivals and resulting AoI evolution.}\vspace{-0.2in}
\end{figure}

A status update is present at time $t$ when $S_t=1$ and it is received when $D_t=1$; otherwise, it is discarded. An example is shown in Figure \ref{fig:model11} to illustrate the AoI evolution for a node applying a power down scheme. $t_i$ denotes arrival times. Updates arriving at $t_1$ and $t_2$ find the node active; hence, it is received. Then the node powers down, misses update at $t_3$ and can capture the one at $t_4$. Similarly, update at $t_5$ is missed due to power down state and update at $t_6$ is received.  

Sequential decisions are performed based on available information. When $D_t=0$ is decided, communication circuts \textit{partially} or \textit{fully} power down, which we elaborate next: 

\noindent \underline{Partial power down (P) scheme:} Radio circuits are powered down while event detection circuits are still active. Device is aware of existence of an incoming update and it uses this information to wake up and receive the status update. 

\noindent \underline{Full power down (F) scheme:} Circuits are powered down so that no information as to the existence of an update is available to trigger wake up. 

Our framework does not account for different energy requirements in these two states. Device consumes zero energy in both (P) and (F) schemes. In practice, devices are designed to jump from (P) to (F) after some time; still, we treat these two extreme cases separately. We assume the decisions are taken by using either (P) of (F) scheme at all times.

The two different power down states yield two different cases of information structures available for decision making. In particular, if the device has the capability to partially power down, at the current time $t$, it has the knowledge of $S_t$ as well as $E_t$ causally (all past and current update arrivals and energy harvests including time $t$). It decides to set $D_t=1$ only when an update is available, i.e., when $S_t=1$. For all $t$ with $S_t=0$, it sets $D_t=0$ to power down. Still, note that due to energy harvesting constraints, the device has to miss some updates by setting $D_t=0$ even when $S_t=1$ in the partial power down. In the full power down, on the other hand, the device is not aware of the updates arriving during the OFF periods. Specifically, $D_t$ is determined by strictly causal knowledge of $S_t$ (all past status updates that have occurred strictly before $t$ during active periods only) as well as causal knowledge of energy arrivals $E_t$ including time $t$. Therefore, in the full power down scheme, it is not possible to use $S_t$ in the decision to power down. We will see that this lack of information causes additional energy cost paid for capturing potential update arrivals and, hence, a compromise in the age of information performance.

\subsection{Age-Threshold Based Policies}

In the remaining sections, our emphasis will be on age-threshold based policies for deciding when to power down in both (P) and (F) cases. Such policies dictate at each time $t$ to set $D_t=1$ only when there is energy in the battery and $\Delta(t) \geq \tau_{th}$, where $\tau_{th}$ is the optimal threshold. Our goal is to understand the effect of such thresholding for the benefit of information freshness. The literature on information freshness analysis in resource constrained nodes provides abundant evidence for usefulness of such policies. For example, \cite{wu2017optimal} proves optimality of threshold based updating in a \textit{generate-at-will} based energy harvesting system in continuous time when $B_{max}=1$ under Poisson energy arrivals. Same work also shows optimality of best effort updating for $B_{max}=\infty$ under no transmission delay. Reference \cite{yates2015lazy} explicitly shows that putting an age-threshold before updating is optimal for $B_{max}=\infty$ when transmission time is not neglected. See also \cite{ceran2019average} for the case of hybrid ARQ under resource constraints. 

In the following sections, in the spirit of the work \cite{wu2017optimal}, we particularly focus on age-threshold based active packet drop control for $B_{max}=1$ and $B_{max}=\infty$. The major difference between our work and \cite{wu2017optimal} is that we focus on the receiver side energy management rather than the transmitter side.

\section{Active Packet Drop Control for $B_{max}=1$}
\label{sec2}

In this section, we consider a node with a single unit battery. In this case, a renewal interval starts with a successful reception of an update by setting $D_t=1$ when $S_t=1$, which causes battery energy to be zero. The next renewal interval $T_n$ is larger than or equal to the time spent to wait for the next energy to arrive. Recall that in our model, the arriving energy is used after it is saved in the battery. Denote the length of time interval between two successive energy arrivals as $I_A \in \{1,2,\cdots\}$, which is geometrically distributed with parameter $q$. Then, $T_n \geq I_{A_n}$ where $I_{A_n}$ is the next energy arrival interval. Then, an update packet is waited and the problem is to decide when to turn the device on. 

Our focus is on \textit{stationary policies} where decisions to turn on ($D_t=1$) are taken based only on the instantaneous value of age or the realization of $I_A=i_A$ in this case\footnote{In our future work, we will address optimality of such policies; still, we keep this issue out of discussion for now.}. The resulting compact form of the achievable average AoI is $\frac{\mathbb{E}[T^2]}{2\mathbb{E}[T]}$, where $T$ is a renewal interval. In the following, we evaluate $T$ for partial and full power down schemes. 

\subsection{Partial Power Down}

In the partial power down scheme, $S_t$ is causally available, and $D_t=0$ whenever there is no status update ($S_t=0$). Therefore, we can express a renewal interval in this case as $T = X(I_A) + I_B$ ,where $X(I_A) \geq I_A$ represents the time at which the decision to receive the next incoming status update is taken ,and $I_B$ is the time for next update to arrive which is geometrically distributed (starting at 0) with success rate $\lambda$.
\begin{align*}
\mathbb{E}[T^2]&=\mathbb{E}[X^2(I_A) + 2X(I_A)I_B + I_B^2], \\
&=\mathbb{E}[X^2(I_A)] + \frac{2(1-\lambda)}{\lambda}\mathbb{E}[X(I_A)] + \frac{(2-\lambda)(1-\lambda)}{\lambda^2},
\end{align*}
and $\mathbb{E}[T]=\mathbb{E}[X(I_A) + I_B]=\mathbb{E}[X(I_A)] + \frac{1-\lambda}{\lambda}$. Minimizing average AoI over $X(I_A) \geq I_A$ functions requires an approach similar to those provided around \cite[Equation (57)]{wu2017optimal}. In particular, we parametrize the problem by forming
\begin{align*}
\hspace{-0.05in}p(\gamma)=\mathbb{E}[X^2(I_A)] + \frac{2(1-\lambda)}{\lambda}\mathbb{E}[X(I_A)] - \gamma \mathbb{E}[X(I_A)] + K(\gamma).
\end{align*}
where $K(\gamma)$ represents additional terms independent of $X(.)$, and the equivalent problem is $\min\{\gamma: \gamma \geq \frac{2(1-\lambda)}{\lambda}, p(\gamma)=0\}$. Constructing the Lagrangian function
\begin{align*}
&L(X(.),\gamma,\mu(.))= \\ & \ \sum_{i_A=1}^{\infty} \left( X^2(i_A) + (\frac{2(1-\lambda)}{\lambda} - \gamma) X(i_A) + K(\gamma) \right)m_{i_A} \\ & - \sum_{i_A=1}^{\infty} \mu(i_A) (X(i_A) - i_A) m_{i_A},
\end{align*}
where $m_{i_A}=q(1-q)^{i_A-1}$ is the probability mass function of $I_A$. We observe that the derivatives with respect to $X(.)$ and $\mu(.)$ have the same forms as those in \cite{wu2017optimal}. As a result, we conclude that $X(I_A)$ must be a threshold based function, i.e., $X(I_A)=\tau_{th}$ if $I_A < \tau_{th}$ or else $X(I_A)=I_A$ for some integer valued $\tau_{th}\geq 0$. We get $\mathbb{E}[T^2] = \frac{(2-\lambda)(1-\lambda)}{\lambda^2} + \tau_{th}^2 + 2\frac{\tau_{th}(1-\lambda)}{\lambda} + (2\frac{\tau_{th}}{q} + \frac{2(1-\lambda)}{q\lambda} + \frac{2-q}{q^2})(1-q)^{\tau_{th}}$ and $\mathbb{E}[T] = \frac{(1-q)^{\tau_{th}}}{q} + \frac{1-\lambda}{\lambda} + \tau_{th}$, which yields the desired closed form average AoI expression, which we minimize over $\tau_{th} \in \mathbb{Z}_{\geq 0}$.

\subsection{Full Power Down}

In the full power down scheme, $S_t$ is strictly causally available and is not available when $D_t=0$. Thus, the receiver has to explore if a status update is present. In this case, an age-threshold based scheme has a renewal interval of the form $T = X(I_{A_1}) + \sum_{k=2}^{N} I_{A_k}$ where $X(I_{A_1}) \geq I_{A_1}$ represents the time at which the decision to set $D_{t}=1$ is taken after the first energy arrival, remaining terms represent the next energy arrival intervals provided that there is no update at time $t$, and $N$ is the number of trials until a status update is captured. As a rule, when $N=1$ the second term $\sum_{k=2}^{N} I_{A_k}$ is zero. Note that $N$ is geometrically distributed with success probability $\lambda$. It is worthwhile to observe that the receiver applies a potential waiting period only after the first energy arrival and the transmissions are done with zero-wait for the rest of the energy arrivals. This is due to the fact that the receiver can make sure that the age of the received update packet is larger than a threshold level with probability one by using such a scheme. We have 
\begin{align*}
&\mathbb{E}[T^2]=\mathbb{E}[X^2(I_{A_1})] + 2(1-\lambda)\mathbb{E}[X(I_{A_1})]\sum_{k=2}^{\tilde{N}} \mathbb{E}[I_{A_k}] \\ & \ \ \  + (1-\lambda)\mathbb{E}[(\sum_{k=2}^{\tilde{N}} I_{A_k} )^2], \\
&=\mathbb{E}[X^2(I_{A_1})] + \frac{2(1-\lambda)}{q\lambda}\mathbb{E}[X(I_{A_1})] + \frac{(2-\lambda q)(1-\lambda)}{q^2 \lambda^2},
\end{align*}
where $\tilde{N}$ represents $N$ conditioned on $N>1$ and expressions follow from memoryless property of geometric distribution, Wald's identity and independence of arrivals. We also have
 \[ 
 \mathbb{E}[T]=\mathbb{E}[X(I_{A_1}) + \sum_{k=2}^{N} I_{A_k}]=\mathbb{E}[X(I_A)] + \frac{(1-\lambda)}{q\lambda}.
 \] 
By evaluating average AoI $\frac{\mathbb{E}[T^2]}{2\mathbb{E}[T]}$ and optimizing over all $X(I_{A_1})$ functions, we can conclude by following similar steps to those in the partial power down scheme that $X(I_{A_1})$ must be a threshold based function, i.e., $X(I_{A_1})=\tau_{th}$ if $I_{A_1} < \tau_{th}$ or else $X(I_{A_1})=I_{A_1}$ for some integer $\tau_{th}\geq 0$. By plugging in the expressions $\mathbb{E}[X(I_{A_1})] = \frac{(1-q)^{\tau_{th}}}{q} + \tau_{th}$ and $\mathbb{E}[X^2(I_{A_1})]=\tau_{th}^2 + (2\frac{\tau_{th}}{q} + \frac{2-q}{q^2})(1-q)^{\tau_{th}}$, we get an AoI expression in terms of the integer valued threshold $\tau_{th}$ which we numerically analyze later.

\section{Active Packet Drop Control for $B_{max}=\infty$}
\label{sec1}

In this section, we consider the other extreme case when the battery size is unlimited. In this case, no incoming energy is lost due to battery overflow. Note that if $q>\lambda$ (energy arrival rate larger than the update arrival rate), then we can set $D_t=1$ for all $t$ and it is trivially feasible and optimal in view of unlimited energy buffer present to save energy. Hence, we assume $q < \lambda$ in the rest. For both partial and full power down schemes, we next propose age-threshold based schemes and evaluate their age performances.

\subsection{Partial Power Down}

For partial power down with unlimited battery, an age-threshold based policy has the following structure: Once a new status update is received, immediately turn off the system by setting $D_t=0$ and wait at least $\tau_{th}$ slots before setting back to $D_t=1$. After $\tau_{th}$ slots, set $D_t=1$ in the first slot with $S_t=1$. This scheme yields the following renewal interval:
\[ T = \tau_{th} + I_B, \]
and for this interval we have $\mathbb{E}[T]=\tau_{th} + \frac{1}{\lambda}$ and $\mathbb{E}[T^2]=\tau_{th}^2 + 2\frac{\tau_{th}}{\lambda} + \frac{2-\lambda}{\lambda^2}$. Due to the unlimited battery size, the energy causality constraints \cite{yates2015lazy,ozel2012achieving} can be expressed as a single constraint on the renewal interval as \[\mathbb{E}[T] \geq \frac{1}{q}, \] which translates into a constraint on the threshold level $\tau_{th} \geq \frac{1}{q} - \frac{1}{\lambda}$. Finally, we observe that the age expression $\frac{\mathbb{E}[T^2]}{\mathbb{E}[T]} = \frac{1}{\lambda} + \tau_{th} - \frac{1-\lambda}{\lambda^2 \tau_{th} + \lambda}$ is monotone increasing with $\tau_{th}$. Therefore, we have to select $\tau_{th}$ as small as possible. 

Due to integer values taken by the variables $T$ and $I_{B}$, we modify our scheme as follows: For a given $\tau_{th}$, take any rational approximation of threshold $\tau_{th}=\frac{f_N}{f_D}$ with $f_N,f_D \in \mathbb{Z}_{>0}$ coprime numbers. Design each renewal interval so as to include $f_D$ status update receptions. Among them apply $\tau_{th}^{(1)}=\lfloor \tau_{th} \rfloor$ for $| f_N - \lceil \tau_{th} \rceil f_D |$ status updates and apply $\tau_{th}^{(2)}=\lceil \tau_{th} \rceil$ for $|f_N - \lfloor \tau_{th} \rfloor f_D|$ status updates. This achieves a long-term average AoI level
\[\frac{| f_N - \lceil \tau_{th} \rceil f_D | g(\tau_{th}^{(1)}) +  |f_N - \lfloor \tau_{th} \rfloor f_D|g(\tau_{th}^{(2)})}{2f_N + \frac{2f_D}{\lambda}}, \]
where $g(\tau) = \tau^2 + 2\frac{\tau}{\lambda} + \frac{2-\lambda}{\lambda^2}$. This method enables us to approximate the term $\mathbb{E}[T^2]$ for any given real valued threshold $\tau$ by a linear combination of $g(\lfloor \tau_{th} \rfloor)$ and $g(\lceil \tau_{th} \rceil)$ which is a compromise that has to be made due to integer constraint on the variables. In particular, the achievable average AoI is
\[\mathbb{E}[\Delta] = \frac{\alpha g(\lfloor \tau_{th} \rfloor) + (1-\alpha)g(\lceil \tau_{th} \rceil) }{2\tau_{th} + \frac{2}{\lambda}}, \] where $\alpha \lfloor \tau_{th} \rfloor + (1-\alpha) \lceil \tau_{th} \rceil = \tau_{th}$. We observe that even after this compromise, the best value of $\tau_{th}$ is still $\frac{1}{q} - \frac{1}{\lambda}$. 

\subsection{Full Power Down}

For full power down with unlimited battery, we apply the same threshold-based scheme as that in the partial power down case. In particular, we immediately turn off the system by setting $D_t=0$ once a new status update is received and wait at least $\tau_{th}$ slots before setting back to $D_t=1$. Since the knowledge of $S_t$ is available only when $D_t=1$ in the full power down scheme, the receiver does not have the option to wait for the next status update arrival. Instead, the receiver sets $D_t=1$ after $\tau_{th}$ slots and keeps active until it captures a new status update. In this case, a renewal interval is once again $T = \tau_{th} + I_B$. Still, there is a subtle difference in the full power down scheme: The receiver is turned on immediately after $\tau_{th}$ and remains during $I_B$ time units until the next status update arrives. Hence, in this case, the rate of power consumption is $\frac{\mathbb{E}[I_B]}{\tau_{th} + \mathbb{E}[I_B]}$ with $\mathbb{E}[I_B]=\frac{1}{\lambda}$ and the power constraint is $\tau_{th} \geq \frac{1}{\lambda}(\frac{1}{q} - 1)$. After this point, we apply identical steps to the partial power down and use the achievable scheme which renders the best selection of threshold is $\tau_{th} = \frac{1}{\lambda}(\frac{1}{q} - 1)$.

\section{Numerical Results}

\subsection{The Case of $B_{max}=0$}

One of the benchmarks to compare age of information performance is the extreme case of no battery. In this case, the receiver can update the incoming packet only if $E_i=1$; otherwise, its power is down. A renewal interval $T$ is geometrically distributed with success probability $q\lambda$. Then, we have the average AoI
\begin{align*}
\Delta = \frac{1}{2} \frac{\mathbb{E}[T^2]}{\mathbb{E}[T]} = \frac{2-q\lambda}{2q\lambda}.
\end{align*}

\subsection{Scheme in \cite{farazi2018age,farazi2018average}}

In earlier closely related work presented in \cite{farazi2018age,farazi2018average}, an energy harvesting node receiving updates from a single information source in continuous time is studied\footnote{To match the problem in \cite{farazi2018age,farazi2018average} to our problem, we set the transmission times to zero. Additionally, our setting is in discrete time.}. In these works, the node is oblivious to energy expended for update packet reception to keep the circuitry active and it ``always accepts" incoming packets if the battery energy is larger than zero. This ``always accept" scheme in \cite{farazi2018age, farazi2018average} is a benchmark for our work. We will make this comparison especially in (P) power down scheme; that is, the node sets $D_t=0$ whenever there is no update $S_t=0$ and always sets $D_t=1$ when $S_t=1$ if there is sufficient energy in the battery.   

For $B_{max}=1$, this scheme is equivalent to setting the threshold $\tau_{th}=0$ in our age-threshold based scheme under partial power down. In this case, a renewal interval is composed of waiting for one unit of energy to arrive and waiting for a status update to arrive (starting from 0 for the latter waiting period). Average AoI is:
\begin{align*}
\Delta = \frac{1}{2}\frac{\frac{(2-\lambda)(1-\lambda)}{\lambda^2} + \frac{2-q}{q^2} + \frac{2(1-\lambda)}{\lambda q}}{\frac{1}{q} + \frac{1-\lambda}{\lambda}}.
\end{align*}

For $B_{max}=\infty$, we note that the resulting battery energy queue is a discrete time Geo/Geo/1 queue with rate $q$ arrivals and rate $\lambda$ departures. Recall that $q < \lambda$ is a standing assumption. Using \cite[Lemma 1]{kosta2019age}, we have the stationary probability of having zero energy in the battery $B_t=0$ \[ \pi_0 = 1 - q - \frac{q(1-\lambda)}{\lambda}, \ \pi_1 = \frac{q}{\lambda (1-q)} \pi_0. \] 
A renewal interval starts by a status update reception, which happens only if a status update arrives when the battery is in $1$ or larger states. Once an update is received successfully, battery level drops by 1. Packet is discarded if battery is empty.  

If an arriving status update finds the battery in $B_{t-1}=1$ at time $t-1$, battery level drops to zero and the next renewal interval is $T_n = I_B + I_A$ where $I_A, I_B$ are geometric with success probability $q$ and $\lambda$, respectively. These variables represent the time waited for a new energy arrival and then a status update arrival (starting from 0 for the latter wait period), respectively. If an arriving status update packet finds the battery in $B_{t-1}=2$ or larger state at time $t-1$, the next renewal interval is $T_n = I_B$; that is, the time waited for a new status update arrival (starting from 1) is counted only since sufficient energy is present. Due to BASTA property of Bernoulli packet arrivals, an arriving status update packet finds the battery in $B_t=0$ and $B_t=1$ states in $\pi_0$ and $\pi_1$ portions of times, respectively. Collecting these results together, we obtain average AoI in closed form:
\begin{align*}
\Delta = \frac{1}{2} \frac{(1-\pi_1-\pi_0) \frac{2-\lambda}{\lambda^2} + \pi_1 (\frac{(2-\lambda)(1-\lambda)}{\lambda^2} + \frac{2-q}{q^2} + \frac{2(1-\lambda)}{\lambda q})}{(1-\pi_1 - \pi_0) \frac{1}{\lambda} + \pi_1 (\frac{1}{q}+\frac{1-\lambda}{\lambda})}.
\end{align*}

\subsection{Comparison of Schemes}

\begin{figure}[t]
\includegraphics[width=1.04\linewidth]{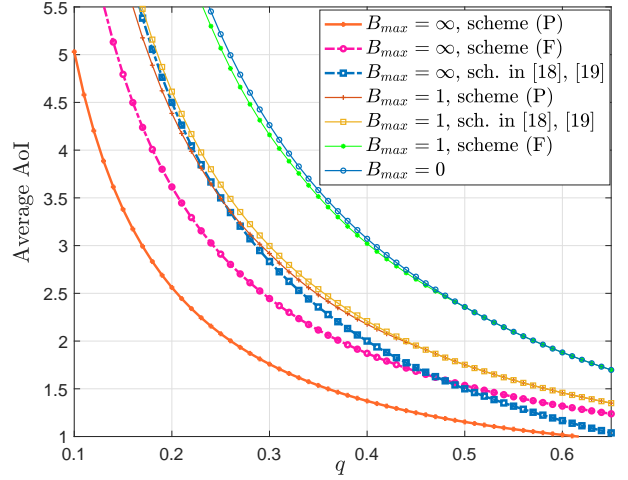}\vspace{-0.15in}
\caption{\label{fig:model} Average AoI versus energy arrival rate for status update rate $\lambda=0.7$.}\vspace{-0.15in}
\end{figure}

In Figure \ref{fig:model}, we compare optimal average AoI versus energy arrival probability $q$ plots for $\lambda=0.7$ under partial (P) and full (F) power down states. The comparisons in Figure \ref{fig:model} with respect to \cite{farazi2018age, farazi2018average} and the other benchmark scheme for $B_{max}=0$ show clearly that an age-threshold based energy management at the receiver yields significant AoI reduction especially with large battery sizes.  

The degree of this reduction varies in partial (P) and full (F) power down states and for different battery sizes. We particularly observe partial (P) power down enables good energy saving for AoI minimization in large battery case. It is also remarkable that a full (F) power down with one unit battery can perform almost as good as having zero battery due to poor energy management, since the receiver is unaware of incoming updates in OFF periods. Also interesting is the comparison between full power down and always accept scheme in \cite{farazi2018age, farazi2018average} especially for $B_{max}=\infty$. This comparison shows that even a poor energy management oblivious to status update arrival times could outperform ``always accept" scheme.

\begin{figure}[t]
\hspace{-0.1in}\includegraphics[width=1.07\linewidth]{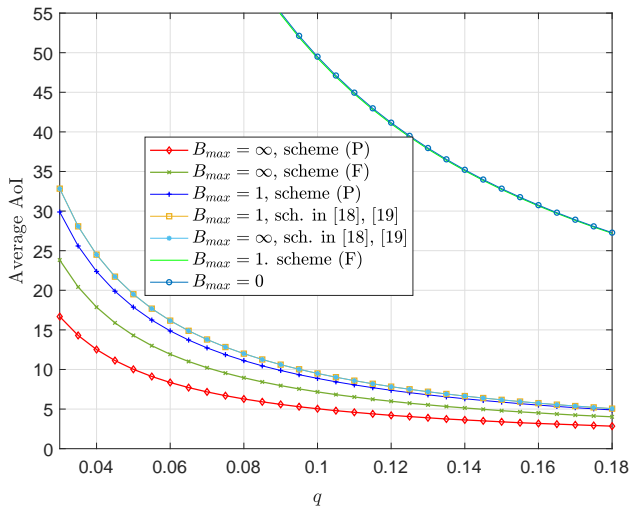}\vspace{-0.1in}
\caption{\label{fig:model2} Average AoI versus energy arrival rate for status update rate $\lambda=0.2$.}\vspace{-0.2in}
\end{figure}

In Figure \ref{fig:model2}, we observe the comparison of different schemes considered for $\lambda=0.2$. In this case, status update arrival rate is significantly smaller and there is a large average AoI gap between $B_{max}=0$ and $B_{max}=1$ with scheme (P), showing the effectiveness of a single battery used along with the capability to wake up and receive a packet while turning off other times. In contrast, scheme (F) with $B_{max}=1$ has no better performance than $B_{max}=0$ (both coincide in the figure). We observe similarly that the scheme in \cite{farazi2018age,farazi2018average} for $B_{max}=1$ and $B_{max}=\infty$ coincide in AoI performance, making it indistinguishable in the figure. This is due to small $\lambda$; hence, just a single battery is sufficient to capture incoming packets to the extent possible. It is also worthwhile to see that (P) scheme brings improvement in average AoI due to its superior energy management capability. As a final remark, once we attempt to obtain the plots for $B_{max}=\infty$ using real valued thresholds $\tau_{th}$ instead of the modified scheme using integer values, we observe in all cases that there are very minor differences and the plots essentially coincide.

\section{Conclusion}

This paper considers age-threshold based active status update packet drop control for an energy harvesting sensor node receiving updates from a single information source. Node decides to turn communication circuitry ON and OFF while harvesting energy throughout. The objective is to keep freshness of information at the sensor node to the extent possible subject to energy harvesting constraints. Our results show that age-threshold based schemes to determine ON-OFF intervals enable significant average AoI improvements for $B_{max}=1$ and $B_{max}=\infty$, giving reason for future research on age-thresholds in optimizing ON-OFF intervals for information freshness in the presence of receiving costs.

\end{document}